\documentclass[a4paper,12pt]{article}

\usepackage[utf8x]{inputenc}
\usepackage[T1]{fontenc}
\usepackage[colorlinks=true,linkcolor=black,citecolor=black,urlcolor=blue]{hyperref}
\usepackage{geometry}
\usepackage{titlesec}
\titleformat{\section}
{\bfseries\uppercase}{\thesection.}{1em}{}
\titleformat{\subsection}
{\bfseries}{\thesection.\thesubsection.}{1em}{}

\usepackage{graphicx} 
\usepackage{multirow} 
\usepackage{cite}
\usepackage{subcaption} 
\usepackage{breakurl}
\usepackage{indentfirst}
\usepackage{amsmath, amssymb, amsfonts, bm,stmaryrd}
\usepackage{txfonts}
\usepackage{fourier}
\usepackage{enumitem}
\usepackage{xcolor}
\usepackage{enumitem}

\titlespacing*{\subsection}{0pt}{1.5em}{0.2em}
\titlespacing*{\section}{0pt}{1.5em}{0.2em}

\renewcommand\eqref[1]{Equation~\ref{#1}}

\renewcommand{\thesection}{\arabic{section}}
\renewcommand{\thesubsection}{\arabic{subsection}}

\makeatletter
\renewcommand\@biblabel[1]{#1.}
\makeatother

\newlength{\bibitemsep}
\newlength{\bibparskip}
\let\oldthebibliography\thebibliography
\renewcommand\thebibliography[1]{%
  \oldthebibliography{#1}%
  \setlength{\parskip}{\bibitemsep}%
  \setlength{\itemsep}{\bibparskip}%
}
\newcommand{\YearConf}{2024}

\newcommand{\CopyrightConf}{Permission is granted for the reproduction of a fractional part of this paper published in the Proceedings of INTER-NOISE \YearConf ~ \underline{provided permission is obtained} from the author(s) \underline{and credit is given} to the author(s) and these proceedings.}

\usepackage{fancyhdr}
  \fancypagestyle{firststyle}
{   \fancyhf{}
   \fancyfoot[C]{\scriptsize \CopyrightConf}
}
\begin{document}
\thispagestyle{firststyle}

\begin{center}
	\includegraphics[width=2in]{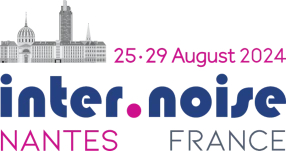}
\end{center}
\vskip.3cm

\begin{flushleft}
\fontsize{16}{20}\selectfont\bfseries
\color{black}Deep learning methods for modeling infrasound transmission loss in the middle atmosphere
\end{flushleft}
\vskip1cm

\renewcommand\baselinestretch{1}
\begin{flushleft}

JANELA CAMEIJO Alice and LE PICHON Alexis\footnote{alice.cameijo@cea.fr, alexis.LE-PICHON@cea.fr}\\
CEA/DAM/DIF\\
F-91297, Arpajon, France\\

\vskip.5cm
AKNINE Samir\footnote{samir.aknine@univ-lyon1.fr}\\
LIRIS, Université Lyon 1\\
F-69130, Ecully, France\\

\vskip.5cm
SKLAB Youcef\footnote{youcef.sklab@ird.fr}\\
IRD, Sorbonne Université\\
UMMISCO, F-93143, Bondy, France\\

\vskip.5cm
ARIB Souhila\footnote{souhila.arib@cyu.fr}\\
Laboratoire Thema, CY Cergy Paris université\\
F-95011, Cergy-Pontoise, France\\

\vskip.5cm
BRISSAUD Quentin and Sven Peter NÄSHOLM \footnote{quentin@norsar.no, peter@norsar.no}\\
NORSAR\\
Gunnar Randers vei 15, 2007 Kjeller, Norway\\

\end{flushleft}
\textbf{\centerline{ABSTRACT}}\\
\textit{Accurate modeling of infrasound transmission losses (TLs) is essential to assess the performance of the global International Monitoring System infrasound network. Among existing propagation modeling tools, parabolic equation (PE) method enables TLs to be finely modeled, but its computational cost does not allow exploration of a large parameter space for operational monitoring applications. To reduce computation times, Brissaud et al.\ 2023\cite{brissaud2023predicting} explored the potential of convolutional neural networks trained on a large set of regionally simulated wavefields ($\le$ 1000 km from the source) to predict TLs with negligible computation times compared to PE simulations. However, this method struggles in unfavorable initial wind conditions, especially at high frequencies, and causal issues with winds at large distances from the source affecting ground TLs close to the source. In this study, we have developed an optimized convolutional network designed to minimize prediction errors while predicting TLs from globally simulated combined temperature and wind fields spanning over propagation ranges of 4000 km. Our approach enhances the previously proposed one by implementing key optimizations that improve the overall architecture performance. The implemented model predicts TLs with an average error of 8.6 dB in the whole frequency band (0.1-3.2 Hz) and explored realistic atmospheric scenarios.}

\section{INTRODUCTION}
Most high-energy atmospheric phenomena, whether natural (meteoroids, earthquakes, volcanoes) or human-made (aircraft, chemical or nuclear explosions), generates waves with frequencies under 20 Hz, below the human audible spectrum, called infrasound. These waves can propagate over thousands of kilometers, thanks to their low attenuation and the layered structure of the atmosphere that guides them, allowing the detection of acoustic information on a global scale. Infrasound are permanently recorded by the International Monitoring System (IMS) set up to detect one kiloton equivalent nuclear explosion around the world (Marty et al.\  2019\cite{marty2019ims}) and monitor the compliance of the Comprehensive Nuclear Test-Ban-Treaty (CTBT). Accurate modeling of infrasound transmission loss (TL) is essential to interpret microbarometer measurements, evaluate their detection thresholds and characterise wavefield parameters (direction of arrival, velocities, amplitudes, frequencies) and source informations (ground pressure levels associated to earthquakes, acoustic energy from man-made or volcanic explosions). TLs modeling can also help to better characterise the middle atmosphere (MA, $\sim 15-100$ km) which significantly impact the infrasound propagation. \\

The computational cost of existing numerical propagation modeling tools, such as normal modes or full-waveform simulations (parabolic equations, PEs, Waxler et al.\ 2021\cite{waxler2021chetzer}), does not currently allow the exploration of a wide parameter space (variations in atmospheric states, representation of small-scale variability, frequency and source location) for near-real time TLs predictions; making them unusable within the required CTBT operational framework. Reducing these computation times by neglecting part of the complexity of the propagation phenomenon introduces significant uncertainties in predicted TLs. For example, Le Pichon et al.\  2012\cite{le2012incorporating} proposed an approach relying on heuristic modelling of wave attenuation using a semi-analytical formula mapping wind speeds in the MA to TLs at ground level. However, this method has been optimized on idealized atmospheric models neglecting range-dependent variations in the atmosphere, resulting in large errors for unfavorable initial wind conditions.

Artificial intelligence methods are currently explored by Brissaud et al.\  2023\cite{brissaud2023predicting} in the Norwegian Seismic Array (NORSAR\footnote{\url{https://www.norsar.no}}) institute. Deep learning algorithms are introduced for accurate TLs estimation ($\sim 5$ dB error compared to PEs) with negligible computation times of the order of $0.05$ s. The proposed algorithm exploits convolutional neural networks (CNNs) to map wind fields simulated on a regional scale ($\le 1000$ km) with TLs at ground level. While results are promising, the maximum propagation distance of $1000$ km is limiting for global events of interest monitoring applications or for the study of certain phenomena such as infrasound generated by colliding ocean waves which generally requires more than $2000$ km propagation paths (Vorobeva et al.\  2020\cite{vorobeva2020microbarom}). Furthermore, as IMS infrasound stations are on average $1700$ km apart, it is necessary to estimate TLs over this distance to draw a comprehensive map of the network detection capability.

The aim of this study is to optimize the convolutional network proposed by Brissaud et al.\  2023\cite{brissaud2023predicting} for rapid estimation of the wave field propagated on a global scale ($\le 4000$ km) in realistic atmospheric conditions. This work will pave the way for a near real-time evaluation of the detection thresholds of IMS stations, a quantification of their location accuracy, and the interpretations of signals generated by sources of interest all over the globe. 

First, the author will detail the databases used to feed the optimized CNN, constructed using PEs simulations. Then, the implementation of the proposed architecture will be presented. Finally, a last section will show the obtained results and an example of a first application case.  

\section{Datasets for transmission losses predictions}
\subsection{Realistic atmospheric slices input dataset}
 
TLs greatly depend on the infrasound propagation through the layered structure of the atmosphere. These trajectories are associated with initial atmospheric conditions such as vertical temperature profiles $T(z)$ or zonal $U(z)$ and meridional $V(z)$ wind speeds. Waves generated by a point source on the ground tend to refract to the upper layers of the atmosphere due to the decrease in temperature encountered in the first kilometers. Rays are reflected back to the ground when they meet the significant temperature rise in the lower thermosphere, up to $120$ km altitude. Zonal winds, from west to east, reach speeds around $50$ m/s between $50-70$ km altitude and can cause additional reflections as the temperature rises in the stratosphere ($35-60$ km altitude).

In order to represent a large set of initial atmospheric conditions leading to various trajectories, $36454$ atmospheric slices of $120$ km altitude and $4000$ km distance are collected around the globe, sampled between latitudes -40°/70°, longitudes -150°/165° and years 2010/2020. 

Slices are horizontally discretized with vertical profiles $T(z)$, $U(z)$ and $V(z)$ extracted every 100 km distance. Similary as described in Brissaud et al.\  2023\cite{brissaud2023predicting}, atmospheric parameters are extracted from the ERA5 re-analysis model up to $\sim 80$ km altitude, with a vertical discretization on 137 altitude levels and an horizontal resolution of 1° (Hersbach et al.\  2020\cite{hersbach2020era5}). Above 80 km altitude, limit of the ERA5 model, $U(z)$ and $V(z)$ are extracted from the HWM-14 climatological model (Drob et al.\  2015\cite{drob2015update}) and $T(z)$ from the NRLMSIS-00 empirical model (Picone et al.\  2002\cite{picone2002nrlmsise}). ERA5 and HWM-14 / NRLMSIS-00 data are connected by cubic interpolation between $75-85$ km altitude. 

\begin{figure}[h!]
\begin{center}
  \begin{subfigure}{3.3in}
    \includegraphics[width=\linewidth]{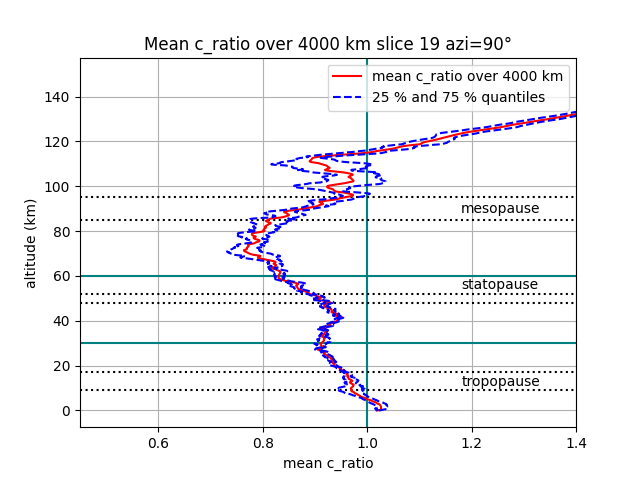}
  \end{subfigure}%
  \begin{subfigure}{3.3in}
    \includegraphics[width=\linewidth]{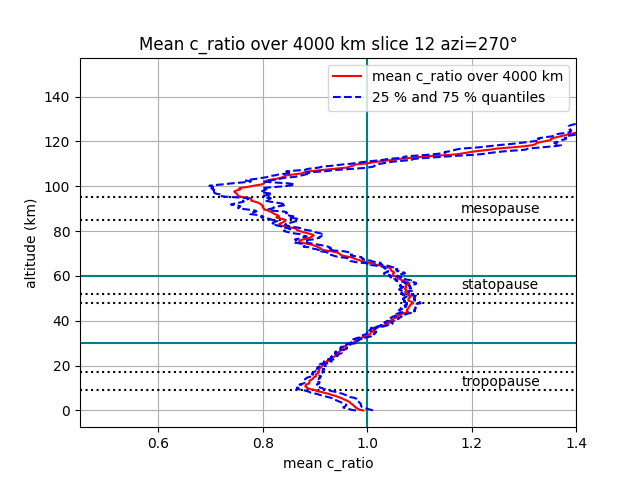}
  \end{subfigure}%
  \caption{Mean $c_\text{ratio}(z)$ over 4000 km and 25 \% and 75 \% quantiles for two scenarios.}
  \label{Fig:1}
\end{center}
\end{figure}

Every 100 km distance, $T(z)$, $U(z)$ and $V(z)$ vertical profiles are combined to compute the effective sound speed ratio $c_\text{ratio}(z)$ defined as :

\begin{eqnarray}
c_\text{ratio(z)} = \frac{c_\text{eff}(z)}{c_{0km}} = \frac{c(z) + v_{0,H}(z)}{c_{0km}},
\label{Eq:1}
\end{eqnarray}

where $c_{0km}$ represents the sound speed at ground level and $c_\text{eff}(z)$ the effective sound speed in a specific atmospheric layer.
$c(z)=c_0 \times \sqrt{T(z)/T_0}$ (m/s) is the adiabatic sound speed depending on temperature profiles, $c_0 = 344$ m/s the speed of sound at 20 °C, $T_0 = 293.13$ K and $v_{0,H}(z) = \sqrt{V(z)^2}$ (m/s) the along-path wind speeds projected in the wave propagation direction. The $c_\text{ratio}(z)$ is computed because it can well explain the trajectories of infrasound as they propagate through the atmosphere. For sources at ground level and at a given altitude $z$, the $c_\text{ratio}(z)$ indicates the probability of the wave being reflected back to the surface. If $c_\text{ratio(z)} \ge 1$, a wave guide will appear and the wave will be reflected to the ground. As wind conditions vary greatly over $4000$ km, atmospheric slices thus constructed are range-dependent with vertical $c_\text{ratio}(z)$ profiles varying in distance. Figure \ref{Fig:1} shows two slices where the mean $c_\text{ratio}(z)$ along the path has been plotted, as well as the 25 \% (q1) and 75 \% (q3) quantiles. The left-hand slice is an example of "upwind" scenario with unfavourable wind conditions for propagation ($c_\text{ratio(z)} \le 1$ between $30-60$ km altitude), leading to the absence of stratospheric guide wave and to a strong attenuation. The right-hand slice is an example of "downwind" situation with favourable wind conditions leading to a stratospheric guide reflecting infrasound back to the ground.

The 1° step used for the ERA5 data extraction is a trade-off between the parameters extraction time and accounting the spatial variability of the propagation medium to be represented (Hersbach et al.\  2020\cite{hersbach2020era5}). However, this resolution is not precise enough to capture small-scale variations in winds and temperatures, such as those due to gravity waves breaking in the troposphere (Chunchuzov et al.\  2019\cite{chunchuzov2019internal}). These fine perturbations are added to the $36454$ slices using a range-dependant spectral model (Gardner et al.\  1993\cite{gardner1993gravity}). Each realisation of this model is superimposed to the original atmospheric specifications and considerably increase the variability and the complexity of the initial conditions, aiming providing a more realistic input dataset.

\subsection{Transmission losses ground-truth dataset}
 
TLs represent the expected outputs of the neural network, i.e. their labels. For each initial atmospheric slice, a TL is associated, calculated over 4000 km in a single dimension (at ground-level only) for a single frequency. The 4000 km distance is crucial to draw a complete detectability map of the IMS network. The single frequency associated to each TL allows to increase the diversity encountered in the label dataset. Given same initial atmospheric conditions, different frequencies lead to different TLs. These frequencies are uniformly sampled between $0.1$ and $3.2$ Hz.

TLs are computed using the Sutherland-Bass attenuation model (Sutherland et al.\  2004\cite{sutherland2004atmospheric}) and by numerically solving the parabolic equations (PEs) of wave propagation in the atmosphere. PEs are complex linear partial differential equations used to find the pressure field $P(z,H)$ in altitude $z$ and in an horizontal plane $H$ originally coming from given atmospheric conditions. Attenuation values are retrieved by computing the modulus of the complex pressure field. 

PEs assume a vertically stratified atmosphere in mean density $\rho_{0}(z)$, sound speed $c(z)$, horizontal wind speeds $v_{0,H}(z)$ and sound pressure $\hat{p}_{A}(x_{H},z,\omega)$ but neglect non-linear propagation and crosswinds effects. These simplifications yield to obtain a Helmholtz convective type equation :

\begin{eqnarray}
\left[\nabla_{H}^2 + \rho_{0}(z) \times \frac{\partial }{\partial z}\left(\frac{1}{\rho_{0}(z)} \times \frac{\partial }{\partial z}\right) + k^2(z)\right] \times \hat{p}_{A}(x_{H},z,\omega) = 0,
\label{Eq:2}
\end{eqnarray}

where $k(z) = \frac{\omega}{c_\text{eff}(z)}$ is the effective wave number and $c_\text{eff}(z)$ the effective sound speed (see Eq. \ref{Eq:1}). The ePaPe PE code used to solve these equations comes from the National Center for Physical Acoustics (NCPA) with a Padé coefficient $M = 7$ (Waxler et al.\  2001\cite{waxler2021chetzer}). This parameter allows the resolution of PEs in the stratosphere and beyond ($z \ge 65$ km altitude). The higher the order of Padé $M$, the broader the simulation of propagation and the more it accounts of refracted rays in the upper layers of the atmosphere.

It is important to note at this point that the expensiveness of PE method, partly due to its accuracy, only appears during the database creation stage. Once done, the proposed neural network will learn on it and then make predictions on new cases almost instantaneously.

\section{Method : convolutional neural network}

Similary to Brissaud et al.\  2023\cite{brissaud2023predicting}, we seek to associate initial atmospheric conditions and TLs at a given frequency using deep learning algorithms called "neural networks". Neural networks are versatile artificial intelligence algorithms designed to address a wide range of problems, including classification, regression, such as curve predictions, and beyond. In "supervised learning", networks work with labelled data and become parametric estimators of functions that perform non-linear mapping between input elements and labels. This mapping results from a learning process during which network parameters (weights and biases) are updated by an optimizer, often using algorithms such as gradient descent, to minimize a loss function. An advantage of such method is the almost instantaneous inference time once the learning process has been completed. This makes neural networks well suited for the operational constraints of the infrasound sources of interest monitoring. 

\begin{figure}[h!]
\begin{center}
  \includegraphics[width=6.5in]{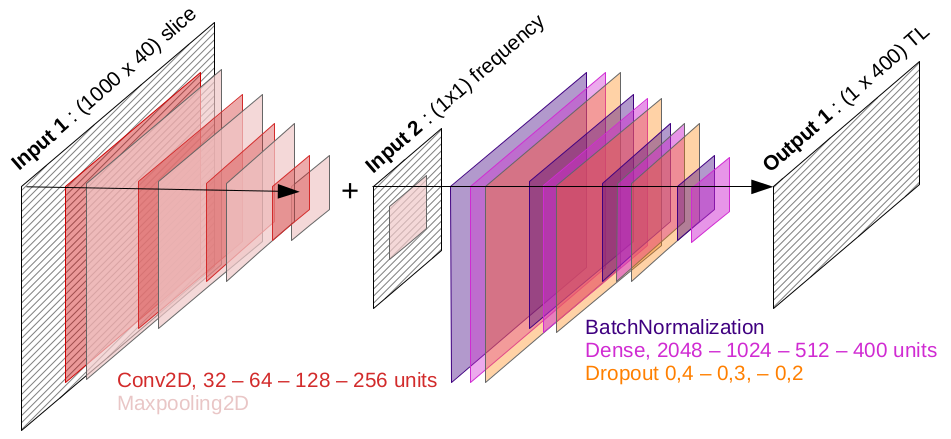}
  \end{center}
  \caption{Optimized 2D-CNN's architecture, containing 7425682 trainable parameters}
  \label{Fig:2}
\end{figure}

Convolutional neural networks are specialized in combining and exploiting local spatial correlations in order to extract meaningful patterns in continuous data like matrices or time-series (Le Cun et al.\  2015\cite{lecun2015deep}). In general, CNNs first learn a new representation of input data by encoding their most relevant features in smaller latent structures and then learn how to associate them with expected outputs. Since inputs are atmospheric slices represented by matrices with a sequential notion of distance, CNNs are well suited to encode them. We thus expect the CNN to effectively catch the logic implied by adjacent range-dependent $c_\text{ratio}(z)$ profiles using its neighbouring features extraction capabilities. The model proposed in this study is an optimization of the 2D-CNN published in Brissaud et al.\  2023\cite{brissaud2023predicting}, which predicted TLs over only $1000$ km. Its architecture is shown in Figure \ref{Fig:2}.

Input data are the atmospheric slices of 120 km altitude and 4000 km distance from which vertically perturbed $c_\text{ratio}(z)$ profiles were extracted every 100 km. Unlike in Brissaud et al.\  2023\cite{brissaud2023predicting}, slices are not downsampled in order to preserve all the spatial variation subtleties, resulting in $(1000 \times 40)$ input matrices. Some of these matrices are used to build the network's training base which is normalized to speed up learning and minimise the risks of explosion/disappearance of the gradient. An operator stores the parameters needed to centre and reduce these data. 

To hierarchically encode the more relevant spatial features, the optimized CNN stacks four 2D convolutional layers with an increasing number of kernels ($32-64-128-256$). These kernels, of constant $(3 \times 3)$ dimensions, contain each 9 parameters that can be updated during learning to better extract local features. Activation functions used in convolutional layers are hyperbolic tangent, expressed as :

\begin{eqnarray}
\tanh(x) = \frac{e^{x} - e^{-x}}{e^{x} + e^{-x}}.
\label{Eq:3}
\end{eqnarray}

They are smooth and differentiable functions that incorporate non-linearity in the network, allowing the CNN to learn complex relationships between varying $c_\text{ratio}(z)$ profiles and TLs at ground level. We use hyperbolic tangent because it transforms inputs $x$ to outputs between -1 and 1, centered in 0, which corresponds to the initial training set normalization process and allows to avoid the exploding gradient problem where some values become arbitrary high during training. Finally, each convolution is followed by a pooling operation which progressively reduce the matrices size and limit the number of parameters to be learned. Filters of dimension $(2 \times 2)$ go through the features map produced by convolution layers with a step of 1, keeping only the maximum value of the filtered pixels. The use of max-pooling instead of average-pooling produced better results for predicting TLs over $4000$ km. At the end, the initial $(1000 \times 40 \times 1)$ atmospheric slices are encoded into $(8 \times 1 \times 256)$ features maps.  

Once the encoding stage is completed, frequencies associated with TLs to be predicted are added. Each latent representation is transformed into column vectors by flattening operations, to which its corresponding frequency is concatenated. Final mappings between extended vectors and their expected TLs are carried out by four fully-connected layers of decreasing width (of respectively $2048$, $1024$, $512$ and $400$ units). To stabilize the mapping process and increase the CNN's robustness, batch normalization operations are performed before each fully-connected layer. They allow to re-center and re-scale mini-batches of data through the network, reducing the variations of data distributions despite the architecture depth. Since batch normalizations keep the average output close to 0 and the standard deviation close to 1, we use rectified linear unit activation functions (ReLU) activation functions in fully-connected layers without risk gradient explosion. Theses functions are linear for values greater than zero and nonlinear as a whole because negative values are always output as zero, ranging outputs between 0 and $+\infty$ with the advantage of reducing the risk of vanishing gradient. Only the last $400$-units-fully-connected layer does not have ReLU as an activation function, but the linear one, which allows to predict TLs at ground level on 400 points (1 value every 10 km distance). 

The quantification of the difference between predicted TLs $\hat{Y}_{i}$ and expected labels $Y_{i}$ is represented by the Mean Square Error (MSE) loss function adapted to regression problems :

\begin{eqnarray}
MSE = \frac{\sum_{i=1}^{N}(Y_i - \hat{Y}_{i})^2}{N},
\label{Eq:4}
\end{eqnarray}

with $N$ the number of samples. The higher the MSE, the more the network weights and biases will be updated to minimise it. Since input slices and labels were normalized before training, the MSE no longer corresponds to a difference in decibels between predicted and expected TLs. The search for the loss function minimum is carried out by the adaptive moment estimation optimizer (Adam) with a learning rate $\eta = 0.0001$ decreasing by a factor of 10 at each iteration from the tenth. The learning process takes place over a maximum of 150 iterations with the possibility of stopping it when the loss function decrease on validation data stagnates beyond 25 iterations. Input data are divided into mini-batches of 32 elements, which prevent the network from over-fitting training samples and no longer being able to generalise on new cases. 

In addition to mini-batches and batch normalization, two other regularisation techniques are finally used to increase CNN's robustness, speedup learning and avoid overfitting. Dropout layers are added after the first three fully-connected layers to randomly deactivate 40 \%, 30 \% and 20 \% of connections between neurons during training and prevent the network from always using the same features to make predictions. All weights are also initialized before the first layer of the network. Similary to Brissaud et al.\  2023\cite{brissaud2023predicting}, we use the uniform Glorot initializer (Glorot et al.\  2010\cite{glorot2010understanding}) which draws samples from a uniform distribution within $\bigg[-\sqrt{\frac{6}{N_\text{in} + N_\text{out}}}, \sqrt{\frac{6}{N_\text{in} + N_\text{out}}}\bigg]$, where $N_\text{in}$ is the number of input units in the weight tensor and $N_\text{out}$ the number of output units. This initializer takes into account the number of parameters in each layer and limits numerical instabilities.

\section{Results}

To evaluate the performances of the proposed CNN, we divide the initial $36454$-slices-dataset in three groups : training, validation and testing sets. Training data are used during learning to progressively update network's weights and biases, validation data are injected at the end of each training iteration to ensure that the learning process goes well and testing data correspond to samples never seen before, used to measure network's generalization capabilities. They respectively represent 70\%, 20\% and 10\% of the initial dataset. We realize a cross-validation procedure with eight independent splits of the dataset in order to train, validate and test the architecture several times and be more confident in its performances. This technique generally results in less biased/optimistic estimates of model skills using limited samples.

\begin{figure}[h!]
\begin{center}
  \includegraphics[width=4.15in]{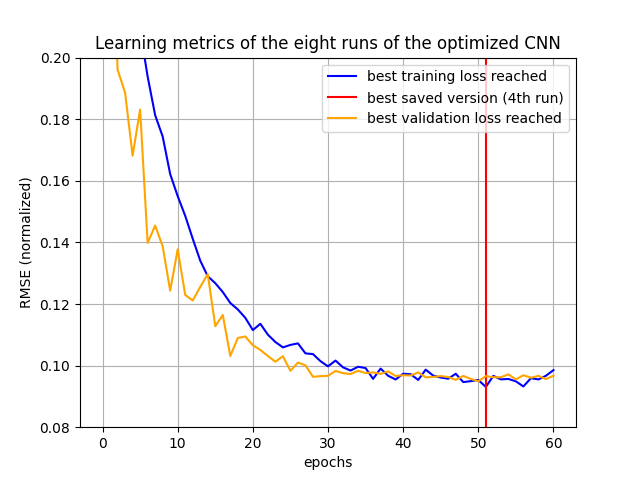}
  \end{center}
  \caption{Training and validation metrics reached by the best network among the eight runs of the optimized CNN.}
  \label{Fig:3}
\end{figure}

Figure \ref{Fig:3} draws training and validation normalized RMSE = $\sqrt(MSE)$ values achieved by the best network among the eight runs of the optimized CNN on eight training and validation sets. Best network reaches around $0.1$ normalized RMSE values in both training and validation after around $40$ iterations. For every run, learning was interrupted when the loss function decrease on validation data stagnated beyond $25$ iterations and last best version's weights and biases were saved in ".h5" files. The red vertical line in Figure \ref{Fig:3} shows the exact iteration at which the best run was last saved. A training loss of 0.0953 and a validation loss of 0.0949 were reached at the $51^{th}$ epoch. Its ".h5" file was reloaded to realize predictions on its corresponding testing set.

\begin{figure}[h!]
\begin{center}
  \begin{subfigure}{3.1in}
    \includegraphics[width=\linewidth]{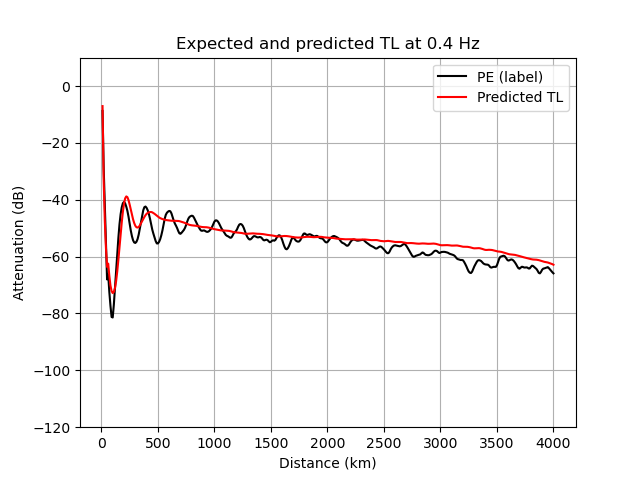}
  \end{subfigure}%
  \begin{subfigure}{3.1in}
    \includegraphics[width=\linewidth]{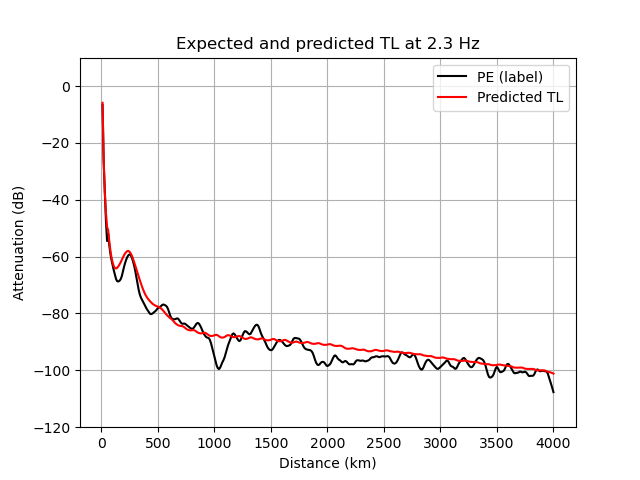}
  \end{subfigure}%
  \vspace{1em}
  \begin{subfigure}{3.1in}
    \includegraphics[width=\linewidth]{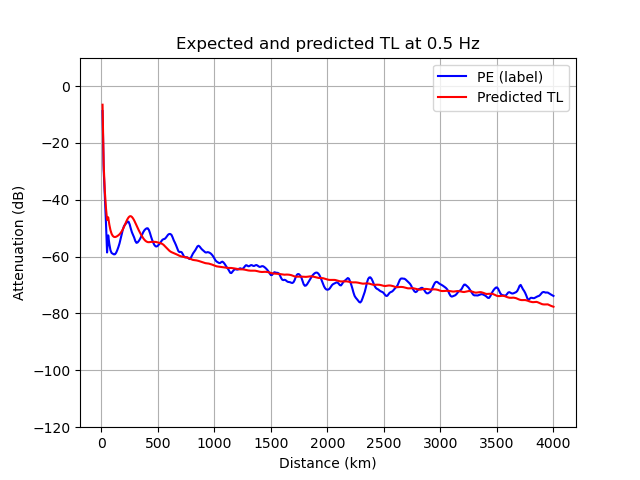}
  \end{subfigure}%
  \begin{subfigure}{3.1in}
    \includegraphics[width=\linewidth]{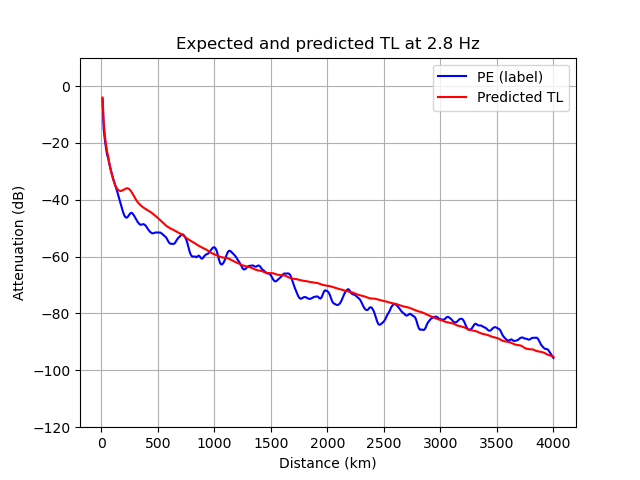}
  \end{subfigure}%
  \vspace{1em}
  \begin{subfigure}{3.1in}
    \includegraphics[width=\linewidth]{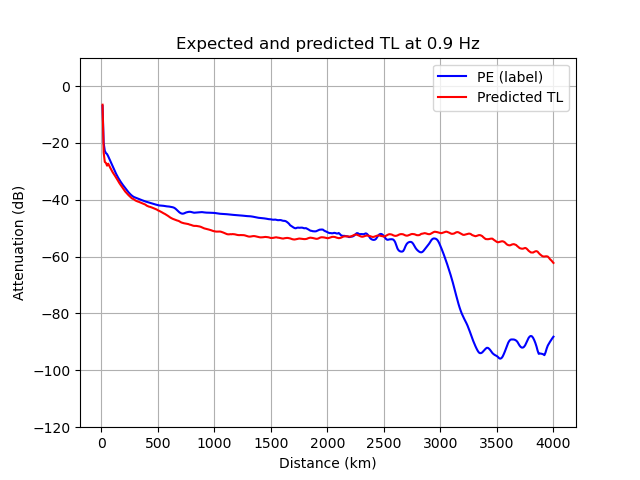}
  \end{subfigure}%
  \begin{subfigure}{3.1in}
    \includegraphics[width=\linewidth]{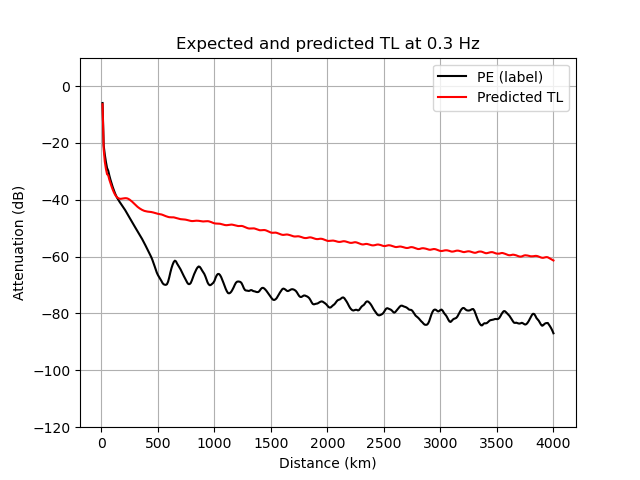}
  \end{subfigure}
  \caption{Examples of predicted and expected TLs (dB) on the best run's testing set.}
  \label{Fig:4}
\end{center}
\end{figure}

Figure \ref{Fig:4} represents six examples of TLs predictions (in red) compared with their expected outputs coming from this testing set. Curves were de-normalized by a process inverse to that described in Section 3 in order to re-expressed attenuations in decibels. Upper images show two well predicted TLs associated with initial favourable wind conditions (downwind cases) with a slight deterioration in performances at high frequency (upper-right image). We explain this by neural networks difficulties in predicting high-frequency features (Rahaman et al.\  2019\cite{rahaman2019spectral}) and by CNNs tendency to smooth local patterns scanned by its kernels like fine spatial variations visible in high-frequencies TLs. Middle images show two other well predicted TLs associated with initial unfavourable wind conditions (upwind cases). The same analysis about decrease in performances at higher frequencies can be done here. Finally, bottom images show two examples of badly predicted TLs.

\begin{figure}[h!]
\begin{center}
  \begin{subfigure}{3.1in}
    \includegraphics[width=\linewidth]{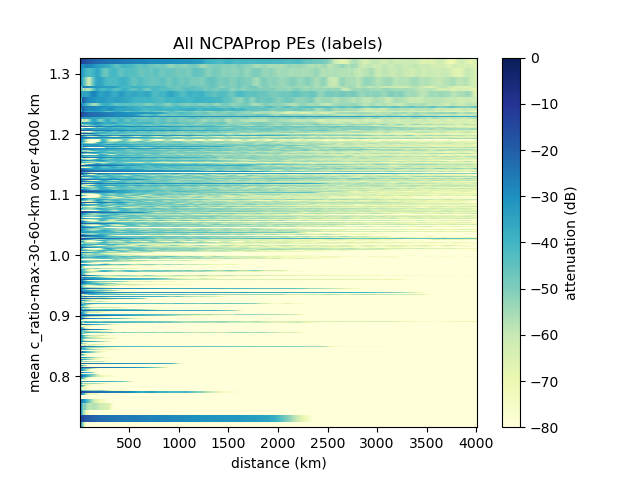}
  \end{subfigure}%
  \begin{subfigure}{3.1in}
    \includegraphics[width=\linewidth]{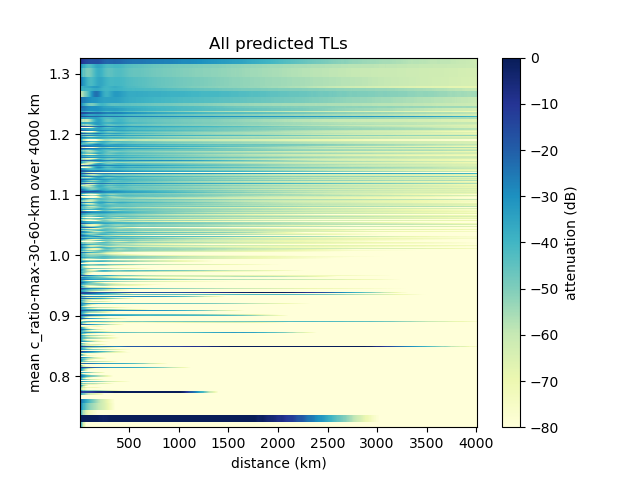}
  \end{subfigure}%
  \vspace{1em}
  \begin{subfigure}{2.9in}
    \includegraphics[width=\linewidth]{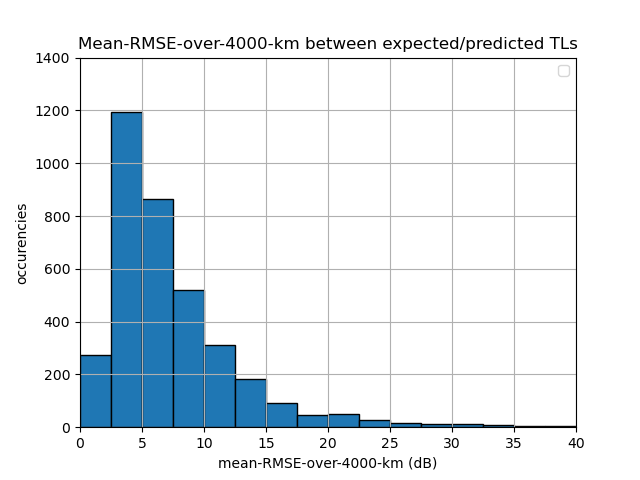}
  \end{subfigure}
  \begin{subfigure}{2.9in}
    \includegraphics[width=\linewidth]{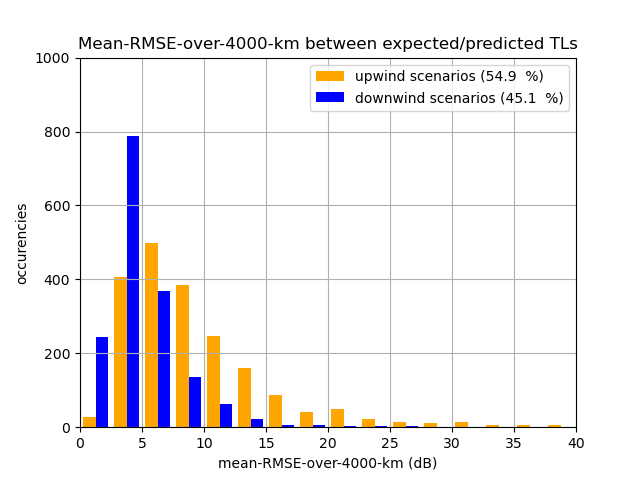}
  \end{subfigure}
  \caption{Global comparisons between 3645 predicted TLs (top-right) and their corresponding labels (top-left) and distribution of the error between them in terms of mean RMSE over 4000 km (bottom).}
  \label{Fig:5}
\end{center}
\end{figure}

\clearpage
A broader evaluation of the best run is carried out in Figure \ref{Fig:5} and help us to better understand badly predicted cases. The $3645$ predictions and associated labels from its testing set have been gathered in two matrices for global comparisons. Scenarios are horizontally arranged, sorted on y-axis by their maximum of mean $c_\text{ratio}(z)$ over 4000 km between $30$ and $60$ km altitude. Cases at the top of each matrix are thus downwind cases, with initial favourable wind conditions and $c_\text{ratio(z)} \ge 1$ in the stratosphere. They are less attenuated over distance than upwind cases gathered at the bottom of the matrices. Analysis of predictions (top-right matrice) compared to their labels (top-left matrice) confirms the CNN tendency to smooth fine spatial variations originally present in PE simulations, particularly at high frequency. More details of prediction errors are presented in the bottom images with two histogramms showing the repartition of cases according to their mean RMSE over 4000 km. Bottom-left figure reveals that the predominant class is the "mean error over 4000 km between 2.5-5 dB" class with $1200$ cases in it among the $3645$ test samples. Remaining classes follow a gaussian distribution, with a wider queue in larger mean error. In addition, bottom-right histogramm shows the impact of downwind or upwind initial conditions on this distribution. We observe a majority of downwind cases (48 \%) gathered in the "mean error over 4000 km between 2.5-5 dB" class against only 21 \% of upwind cases in it, but also 25 \% of upwind cases in the "mean error over 4000 km  between 5-7.5 dB" class and even 18 \% of upwind cases in the "mean error over 4000 km between 7.5-10 dB" class. We explain this by larger effects of fine-scale structures on TLs from upwind cases, compared with downwind cases where a strong stratospheric wave guide represents the major feature to be learn by the network. 

A frequency decomposition of the matrices in Figure \ref{Fig:5} finally allows to evaluate the CNN performance as a function of frequencies. 12 groups are created by gathering the 32 possible frequencies between $0.1-3.2$ Hz according to the operational decomposition used to exploit the IMS network data. Each class is constructed thanks to a logarithmic decomposition following a third octave band scaling with a time windows length ten times longer than the considered period (Brachet et al.\ 2010\cite{brachet2010monitoring}). Table \ref{Tab:1} gathers average mean RMSE over 4000 km and the 25 \% - 75 \% quantile values reach by the best saved optimized CNN for each created frequency group. As explained earlier, performances decrease with increasing frequencies. Despite this, the average error on the best run's test set is 8.6 dB, all frequencies and initial wind conditions considered, which is comparable with the average 5 dB error obtained by Brissaud et al.\ 2023\cite{brissaud2023predicting} predicting TLs over 1000 km. This demonstrates the CNN's capabilities to predict TLs over 4000 km considering realistic atmospheric conditions. 

\begin{table}[h!]
\caption{Error distribution on the test set of the best CNN's run according to the frequency.}
\label{Tab:1}
\begin{center}
\begin{tabular}{c c c c c c c c c} 
 \hline
 \textbf{Frequencies (Hz)} &  \textbf{0.1 - 0.1}& \textbf{0.2 - 0.2} & \textbf{0.3 - 0.3} & \textbf{0.4 - 0.4} & \textbf{0.5 - 0.5}&  \textbf{0.6 - 0.7} \\ [0.5ex] 
  \hline
 number of cases & 54 & 114 & 106 & 119 & 121 & 233\\ 
 \hline
 q1 mean-RMSE-4000-km (dB) & 4.7 & 3.9 & 3.6 & 3.6 & 3.7 & 3.7\\
 \hline
 average mean-RMSE-4000-km (dB)  & 8.9 & 7.3 & 8.0 & 8.1 & 7.5 & 7.8\\ 
 \hline
 q3 mean-RMSE-4000-km (dB) & 9.7 & 9.3 & 9.9 & 10.1 & 9.1 & 10.1\\ [1ex] 
\hline
\textbf{Frequencies (Hz)} & \textbf{0.8 - 0.9} & \textbf{1.0 - 1.1} & \textbf{1.2 - 1.5} & \textbf{1.6 - 1.9}& \textbf{2.0 - 2.4}& \textbf{2.5 - 3.2} \\ [0.5ex] 
 \hline
 number of cases & 250 & 229 & 431 & 496 & 587 & 905\\ 
 \hline
  q1 mean-RMSE-4000-km (dB) & 3.9 & 3.9 & 4.2 & 3.7 & 4.6 & 5.0\\
 \hline
 average mean-RMSE-4000-km (dB) & 8.5 & 8.1 & 9.2 & 8.7 & 10.4 & 11.6\\ 
 \hline
 q3 mean-RMSE-4000-km (dB) & 10.9 & 9.7 & 10.9 & 11.6 & 13.5 & 14.4\\ [1ex] 
 \hline
\end{tabular}
\end{center}
\end{table}

\begin{figure}[h!]
\begin{center}
  \begin{subfigure}{3.1in}
    \includegraphics[width=\linewidth]{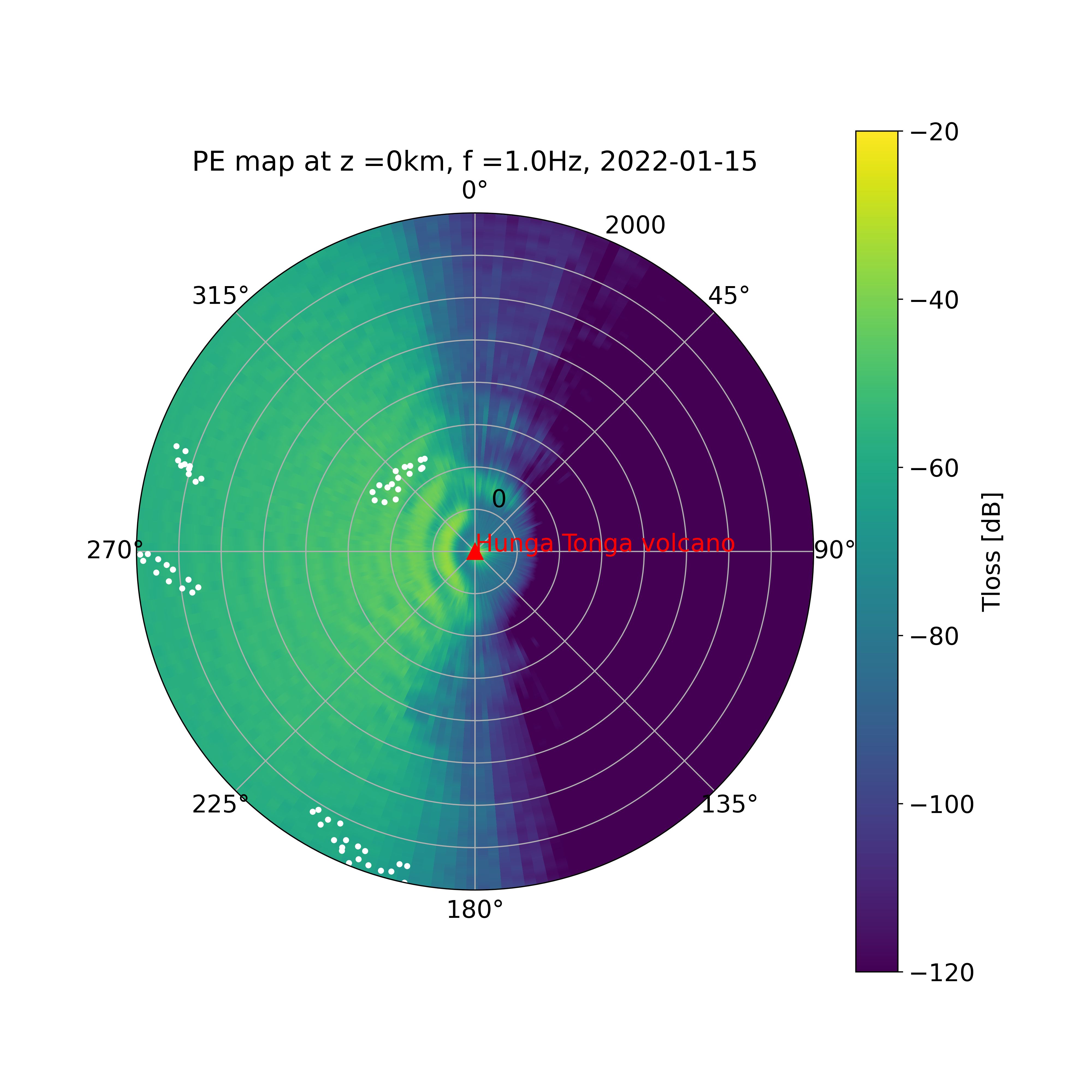}
  \end{subfigure}
  \begin{subfigure}{3.1in}
    \includegraphics[width=\linewidth]{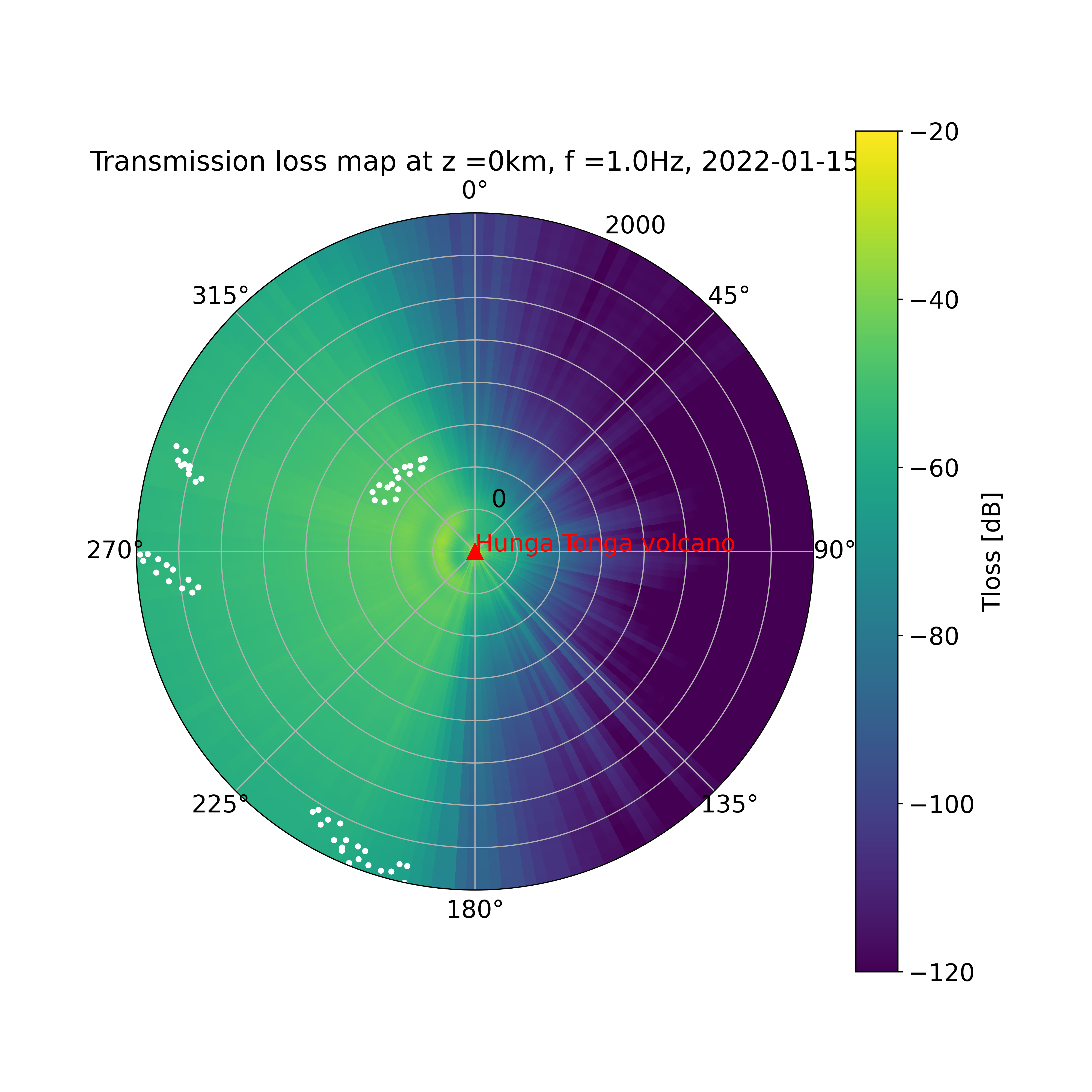}
  \end{subfigure}
  \caption{PE simulations and predicted attenuations maps at 0 km altitude and 1.0 Hz - Hunga Tonga volcano explosion on 2022, January 15.}
  \label{Fig:6}
\end{center}
\end{figure}

A major contribution allowed by this whole work is the possibility to draw global attenuation maps in near-real time around sources of interest. Figure \ref{Fig:6} shows an example of such application associated to the Hunga Tonga volcano explosion on 2022, January 15 (Matoza et al.\  2022\cite{matoza2022atmospheric}). Left image represents the 1.0 Hz PE simulations map at ground-level associated with the January 15 $c_\text{ratio}(z)$ conditions. Right figure gathers for its part the predicted TLs at 1.0 Hz in an attenuation map build using the optimized CNN. PEs simulations and TLs predictions are made over 2000 km from the source, in every directions.  

\section{Conclusion and discussions}

This work shows the potential of deep learning methods to predict infrasound transmission losses over a distance of 4000 km. These estimates can be used to provide a better diagnosis of the impact of multi-scale atmospheric structures on the propagation of infrasound. In addition, the almost instantaneous inference time of neural networks makes them applicable in an operational framework to monitor events of interest on a global scale. Once the number of stations being able to detect an event is determined, its location accuracy can also easily be determined in near-real time. This is a major advantage compared with conventional numerical simulation methods.

The proposed architecture is an optimized version of the 2D-CNN originally developed by Brissaud et al.\  2023\cite{brissaud2023predicting}. It runs using realistic range-dependant atmospheric slices from which vertically perturbed $c_\text{ratio}(z)$ profiles were extracted. This represents an improvement compared to the use of only wind profiles, in that $c\_ratio(z)$ can better explain wave trajectories and improve predictions interpretability. A second major improvement is the increasing prediction distance up to 4000 km, which will allow to draw attenuation maps and complete detection maps of IMS network stations on a global scale.

Perspectives include the use a more realistic data, for example by adding finer atmospheric variability at high altitude using models such as the Whole Atmosphere Community Climate Model (WACCM). We could also imagine a different label set build using normal modes instead of computationally expensive PEs simulations. We could thus compare prediction performances, and even realize a sensitivity study of the network according to its labels. Another specific interest is the TLs predictions for sources in altitude in the full propagation plan (not only at the ground level). A last improvement would be the addition of confidence levels associated with TL predictions. 

Finally, this study could be extended by including more physical laws in learning processes. Neural operators could, for example, be used to solve partial differential equations of wave propagation with a large variability of initial and boundary conditions (Li et al.\  2021\cite{li2020fourier}). 

\section*{Acknowledgements}

We gratefully acknowledge the Defence Innovation Agency for co-financing the thesis that led to the development of this work.

\bibliographystyle{unsrt}
\bibliography{sample} 

\end{document}